\documentclass[epsf,twocolumn,showpacs,preprintnumbers]{revtex4}
\usepackage{graphics}
\usepackage{graphicx}
\usepackage{dcolumn} 
\usepackage{bm}
\usepackage{epsfig}
\begin{document}
\title{Origin of Superconductivity in Boron-doped Diamond}
\author{K.-W. Lee and W. E. Pickett}
\affiliation{Department of Physics, University of California, Davis, CA 95616}
\date{\today}
\pacs{71.20.-b, 71.20.Be, 71.20.Eh, 71.27.+a}
\begin{abstract}

Superconductivity of boron-doped diamond, reported recently at T$_c$=4 K,
is investigated exploiting its electronic and vibrational analogies to
MgB$_2$.  The deformation potential of the hole states arising from the
C-C bond stretch mode is 60\% larger than the 
corresponding quantity in
MgB$_2$ that drives its high T$_c$, 
leading to very large electron-phonon matrix elements.  The 
calculated coupling strength $\lambda \approx$ 0.5 leads to T$_c$ in 
the 5-10 K range and makes phonon coupling the likely mechanism.
Higher doping should increase T$_c$ somewhat, but
effects of three dimensionality primarily on
the density of states 
keep doped diamond from having a T$_c$ closer to
that of MgB$_2$.

\end{abstract}
\maketitle

Discovery of new types of superconducting materials has accelerated since
the discovery of the high temperature superconductors, with a
recent breakthrough being the discovery of 
superconductivity at T$_c$= 40 K in the 
simple (structurally and electronically) compound MgB$_2$.\cite{akimitsu}  
The origin of
its remarkable superconductivity is now understood to arise from charge
carriers doped (in this case, self-doped) into very strongly bonding 
states that in turn respond very sensitively to the bond-stretching vibrational
modes.\cite{jan}
The strong B-B bonds in the graphitic layers of MgB$_2$ make it 
appear near optimal, although graphite itself and diamond are materials that
have even stronger bonds.  Of these two, only diamond has bonding states that
can conceivably become conducting through hole-doping.\cite{jan2}  
The recent report 
by Ekimov {\it et al.}\cite{ekimov} of superconductivity at 4 K in very
heavily boron-doped diamond revives the question of mechanisms in strongly
covalent materials. Confirmation has been provided by 
Takano {\it et al.} who report T$_c$ = 7 K in B-doped diamond 
films.\cite{takano}  

While the study of B as a hole dopant in diamond has a long history,
there have been recent developments due to the ability to dope diamond films
more heavily (to and beyond the 10$^{20}$ cm$^{-3}$ range) than was possible  
previously.  In spite of its growing importance, and unlike the situation
for donors,\cite{bernholc} there has been little
theoretical work on the acceptor state (such as determining its spatial
extent) beyond obtaining the structural and vibrational properties of
the isolated B impurity.\cite{briddon}
An isolated B atom is an acceptor with a binding energy 
of 0.37 eV\cite{thonke} for which effective mass theory is not applicable,
but the behavior of B-doped diamond up to and somewhat beyond the concentration
for insulator-metal (Mott) transition 
$c_M$=2$\times$10$^{20}$ cm$^{-3}$ has been
rather well studied experimentally.\cite{thonke}  
(The B concentration achieved by Ekimov {\it et al.}
is about $c_{sc}$=5$\times$10$^{21}$ cm$^{-3} = 25 c_M$, with a hole carrier density of
nearly the same, and introduces a new regime of metallic diamond that is
yet to be understood.)  At this concentration, B dopants
are on average 5-6~\AA~apart and their donor states, which form an
impurity band already at $c_M = \frac{1}{25}c_{sc},$ broaden considerably 
and overlap the valence band edge.  In
addition, Mamin and Inushima have pointed out\cite{mamin} 
that as the B concentration
increases (and they were not yet thinking of the 10$^{21}$ range) 
many of the donor states become more weakly bound states of B-related
complexes that would encourage formation of broader bands.
Fontaine has analyzed the concentration dependence of the activation 
energy\cite{fontaine} (0.37 eV at low concentration) 
and concluded that it vanishes at
$c = 8\times 10^{20}$ cm$^{-3}$ = $\frac{1}{6} c_{sc}$; for larger 
concentrations the system would be a degenerate metal.

Given these indications of degenerate behavior even below 2.5\% doping,
in this paper we adopt the viewpoint that the majority fraction of the hole
carriers resides in states overlapping the diamond valence band, and
behave primarily as degenerate valence band holes. 
Boeri {\it et al.}\cite{oka} have taken a similar viewpoint, and two
supercell calculations\cite{xiang,blase} have verified this 
degenerate metal picture. 
The distinctly different low-concentration, nonmetallic limit has also 
been suggested.\cite{baskaran}
We investigate the
magnitude and effect of hole-phonon coupling analogously to what has been
found to drive superconductivity in MgB$_2$,
and present evidence that at hole-doping
levels similar to that reported, the hole -- bond-stretch coupling is
surprisingly strong and makes phonon exchange
a prime candidate for the mechanism of pairing.
In the case that such coupling is
strong, it can be verified by spectroscopy of the Raman-active bond stretch
mode.
In fact, Ekimov {\it et al.} report\cite{ekimov} a Raman spectrum in which the 
sharp diamond peak at 1332 cm$^{-1}$ has vanished, leaving spectral weight 
in the 1000-1300 cm$^{-1}$ range.  This behavior is a plausible extrapolation
(considering they are very differently prepared materials)
from more lightly B-doped films in which Ager {\it et al.} observed an initial
weakening and broadening of the 1332 cm$^{-1}$ mode,\cite{ager}
and a transfer of spectral density to peaks in the 940-980 cm$^{-1}$ range
for concentrations $\sim c_{sc}$.  Zhang {\it et al.} reported, for
films with concentration $\sim \frac{1}{3}c_{sc}$ a broad peak at 1200 
cm$^{-3}$ and a very broad feature peaking at 485 cm$^{-3}$.\cite{zhang}

For the sake of definiteness, we consider a hole concentration of
0.025/carbon atom, 10\% less than the B concentration 
($c_{sc}$) determined for the
superconducting diamond films.  At this concentration the hole Fermi 
energy E$_F$ lies 0.61 eV below the valence band maximum, and the diamond Fermi
surfaces consist of three zone-centered ``spheroids.'' The outer one in
particular differs considerably from spherical due to the anisotropy of the
band mass, but effects of anisotropy will be decreased by disorder scattering
and in any case would give only second order corrections to the properties
that we calculate.

The system we consider is thus 2.5\% hole-doped diamond.  The key points
here are (1) the carrier states are the {\it very strongly} covalent bonding
states that make diamond so hard, and (2) these states should be 
sensitively coupled to the bond-stretching mode, which lies at the very
high frequency of 1332 cm$^{-1}$ (0.16 eV) in diamond.  These ingredients
are the same as those prevailing in MgB$_2$.  There are differences, both of a
positive and negative nature.  In MgB$_2$ only two of the nine phonon branches
are bond-stretching, whereas in diamond these comprise three of the six 
branches.  On the other hand, MgB$_2$ is strongly two dimensional in its
important bands ($\sigma$ bands), which means a near-step-function increase
in the density of participating states as doping occurs; the states in 
diamond are three-dimensional and their Fermi level density of state N(0)
increases with doping level
more slowly.

A look at the phonon spectrum of diamond\cite{phonons} reveals that the three
optic modes are the bond stretching ones, and they have little dispersion
so ${\bar \Omega}_{\circ} \approx 0.15$ eV is their common unrenormalized
frequency.
The theory of carrier-phonon coupling and the resulting superconductivity
in such systems is well developed, and the important features in MgB$_2$-like
systems have been laid out explicitly.  The coupling strength $\lambda$ is
given rigorously for an element by
\begin{eqnarray}
\lambda = \frac{\sum_b N_b(0) <I_b^2>}{M<\omega^2>} 
        =\frac{N(0)~I^2_{rms}}{M\omega^2_{\circ}}
\label{first}
\end{eqnarray}
where $N_b(0)$ is the DOS of band $b$, $M$ is the carbon mass, $I_b^2\equiv 
\ll |I_b(k,k')|^2 \gg _{FS}$ is the Fermi surface averaged electron-ion matrix
element squared for band $b$, and
$<\omega^2>$ is an appropriately defined mean square frequency which
will simplify to $\omega^2_{\circ}$, the bond stretch frequency renormalized
by the hole doping.  
The sum over bands $b$ has been displayed explicitly but finally
leads to an rms electron-ion matrix element in the numerator, and the 
Fermi level density of states is N(0)=0.060 states/eV per cell per spin.

Due to non-sphericity and non-parabolicity of the three inequivalent
bands substantial computation would be required to obtain accurate
numbers (and anharmonic and non-adiabatic corrections would change them,
see below).
There are two ways to obtain approximate values in a
pedagogical manner: (1) calculate the $Q$=0 deformation potentials to
obtain the matrix elements for the optic modes, or 
(2) calculate the phonon softening and use
the lattice dynamical result
\begin{eqnarray}
\omega_{Q}^2&=&\Omega_{Q}^2 + 2\Omega_{Q}~Re~\Pi(Q,0)\nonumber \\
\omega^2_{\circ} = \omega_{Q\rightarrow 0}^2 &\rightarrow& 
         \Omega^2_{\circ} -2~\Omega_{\circ}
      ~N(0)~|{\cal M}|^2.
\label{renormalize}
\end{eqnarray}
where $\Pi(Q,\omega)$ is the phonon self energy arising from the doped
holes, and ${\cal M}$ is the electron-phonon matrix element and is determined 
by $I_{rms}$ (see below).
We apply both methods to
obtain estimates of the coupling strength.

The calculations were done with the Wien2k linearized augmented plane
wave code.\cite{wien} 
The basis size was fixed by $R_{mt}K_{max} = 7.0$
with a sphere radius 1.2 for all calculations.
While 110 irreducible $\bf{k}$ points were used for pure diamond,
1156 $\bf{k}$ points in the irreducible wedge for the
B-doped diamond virtual crystal 
calculations,
(nuclear charge $Z=(1-x)Z_C + x Z_B = 5.975$) 
because the Fermi surface
volume had to be sampled properly to account for screening.
Alloy (coherent potential approximation, CPA) calculations 
using the full potential
local orbital code\cite{fplo} give bands as in our virtual crystal model,
the main difference\cite{cpa} being small disorder broadening that would not 
change our conclusions.

\begin{figure}[bt]
\psfig{figure=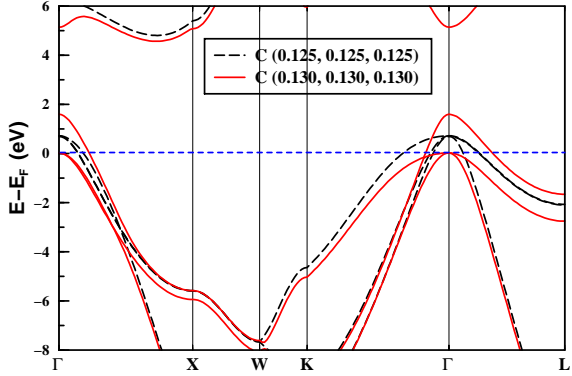,width=5.0cm,angle=-90}
\caption{\label{bands}
(Color online) Virtual crystal bands of 2.5\% B-doped diamond without
(dashed lines) and with (solid lines) a bond stretch phonon frozen in.
The atomic displacement $a\sqrt{3} \Delta x$ = 0.0309~\AA~is just enough 
to transfer all holes to within a single $k$=0 centered Fermi surface.
The horizontal dashed line indicates the Fermi level (aligned for this plot).
X denotes the usual zone boundary point, L designates the zone boundary point
in the $<111>$ direction parallel to the atomic displacements.
}
\end{figure}

The central quantity in Eq. \ref{first} is the matrix element, 
which can be expressed in 
terms of the deformation potential ${\cal D}$; we use the definitions
of Khan and 
Allen\cite{khan} to avoid ambiguity.  
${\cal D}$ is the shift in the hole (valence) band edge 
with respect to the bond stretching motion, whose scale is given by
$u_{\circ} = \sqrt{\hbar/2M\Omega_{\circ}}$ = 0.034~\AA.  
The stretching mode is threefold degenerate and can have any direction of
polarization.  We have chosen the polarization in which atoms move along a
$<111>$ direction.
Under this displacement, the
threefold eigenvalue splits (see Fig. \ref{bands} for the case of doped diamond)
at the rate of $\Delta(\varepsilon_{upper}
-\varepsilon_{lower})_{k=0}/\Delta
d_{bond}$ = 21 eV/\AA, where $d_{bond}$ is the bond length.  Since the 
twofold band splits half as rapidly as the single band (and oppositely)
this leads to the two deformation potentials of magnitude
${\cal D}_1$ = 14 eV/\AA~
for the nondegenerate band and ${\cal D}_2$ = 7 eV/\AA~ for the doublet,
for intrinsic diamond.  The large deformation potential is 60\%
larger than the (already large) analogous one in MgB$_2$.

The results for 2.5\% B-doping are needed for calculation of the coupling
strength, and are shown in Fig. 
\ref{bands}.  They are renormalized by B-doping by 3\% (downward)
from those of intrinsic
diamond, so again we have
${\cal D}_1$ =14 
eV/\AA, ${\cal D}_2$ = 7.0 eV/\AA.
These deformation potentials are undoubtedly the largest yet encountered for any
metallic solid, being directly related to the great bond strength of 
diamond.  Since the three deformation potentials contribute additively to the 
coupling strength, we simplify by using the root mean square value 
$I_{rms}$ = 10 eV/\AA.   The rms electron-phonon matrix element, to be
used below, is ${\cal M} =
        \sqrt{\omega_{\circ}/\Omega_{\circ}}~u_{\circ}~I_{rms}$ =  
0.70 eV; here $\omega_{\circ}$ is the renormalized optic frequency.

Together with the value N(0) = 0.060 states/eV-spin, $M\Omega^2_{\circ}$ =
65 eV/\AA$^2$, and (calculated below) $\omega^2_{\circ} =
0.68 \Omega^2_{\circ}$, we obtain from Eq.
\ref{first} the coupling strength $\lambda =0.55$.  
The coupling is confined to a set of three optic branches which comprise a
narrow peak centered at $\omega_{\circ}$.
The conventional theory, neglecting very minor strong-coupling 
corrections,\cite{allendynes} gives
T$_c = (\omega_{\circ}/1.2)~exp[-1/(\frac{\lambda}{1+\lambda} - \mu^*)]$, 
where $\mu^*$ is an effective
Coulomb repulsion that is uncertain for doped diamond.  Using the 
conventional value $\mu^*$ = 0.15 with $\omega_{\circ}$= 0.128 eV
gives T$_c$= 9 K, gratifyingly (and probably fortuitously)
close to the experimental values of 4 K to 7 K.  
To obtain the initially observed value T$_c$=4 K
would require $\lambda$= 0.48, or alternatively $\mu^* \approx$
0.20, {\it i.e.} relatively small changes.

\begin{figure}[bt]
\psfig{figure=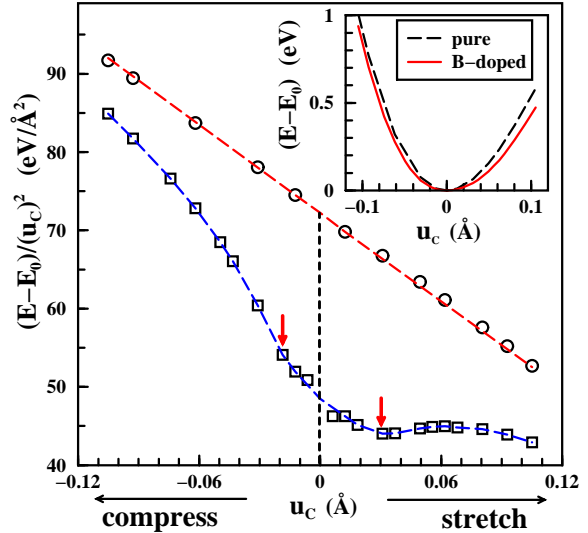,width=7.0cm,angle=-90} 
\caption{\label{energy}
(Color online) Plots of the energy of distortion for the frozen-in bond
stretch mode, for (top straight line) undoped diamond, and (bottom line)
2.5\% B-doped diamond.  The coordinate $u_C$ is for one of the two 
identically displaced atoms.
Pristine diamond follows a quadratic plus lowest-order
anharmonic form $\Delta E(u) = A_2 u^2 + A_3 u^3$ accurately, as indicated
by the straight dashed line.  After doping the 
$\Delta E(u)$ functional form becomes very complex.  The horizontal arrows 
indicate the atomic displacements for which one (or two) Fermi surfaces
disappear.  The inset shows the $\Delta E(u)$ curves themselves. 
}
\end{figure}
 
A less direct way of obtaining the coupling strength, 
but one that (numerically)
includes averaging over bands properly, is to calculate
the renormalized phonon frequency and apply Eq. \ref{renormalize}.  The
calculated change in energy versus atomic displacement, plotted as
$\Delta E(u)/u^2$, is shown in Fig.
\ref{energy}, both for intrinsic and doped diamond.  The difference due to 
doping is
striking.  The result for diamond is simple to understand: the harmonic
$u^2$ term gives $\Omega_{harm}$ = 1308 cm$^{-1}$ similar to literature
values,\cite{phonons,mehl} and the $A_3 u^3$ term quantities its
anharmonicity.

The $\Delta E(u)/u^2$ curve for the doped case (see Fig. \ref{energy})
is much more complex.  The reason is clarified by the red arrows on the plot,
which mark the displacements where some piece(s) of Fermi surface vanishes.  
One of these
values of displacement is also used for the deformation potential plot of
Fig. \ref{bands}, where one shifted band edge is lying exactly at E$_F$.  Boeri
{\it et al.} have described how, at such topological transitions of the
Fermi surface, the 
energy is non-analytic.\cite{boeri}  
In Fig. \ref{energy} one can imagine a straight
line behavior similar to (and nearly parallel to) 
that of diamond for $u_C$ between the two topological transitions, with 
changes of behavior occurring beyond each transition point.  
It is of special relevance
that these positions are roughly at the bond-stretch amplitude, hence they are
physically important.

Returning to the coefficient of the $u^2$ (harmonic) term, for the doped case 
it is 0.68 that of diamond,
giving (in harmonic approximation) $\omega_{\circ}$= 1070 cm$^{-1}$.  
This renormalization of the square of the optic mode frequency (see Eq. 
\ref{renormalize}) by 32\% is a vivid indication of the strong coupling,
even for the case of 2.5\% holes.
Substituting $\omega_{\circ}$ ratio into Eq. \ref{renormalize} allows us to extract the
electron-phonon matrix element ${\cal M}$ = 0.67 eV, within 5\% of the value
obtained from the deformation calculation.  Combined with the deformation
potential result, the predicted coupling strength is
$\lambda = 0.53 \pm 0.03$.  As mentioned above,
this value is quite consistent with the observed critical temperature, but
certainly such good agreement may not be warranted.

Our treatment neglects some complicating features.  The Jahn-Teller splitting
of the isolated B substitutional impurity 
(0.8 cm$^{-1}$ from Fabry-P\'erot spectroscopy\cite{kim})
is three orders of magnitude smaller
than energy differences involved in the bond-stretch mode and
therefore is negligible.  It has been suggested\cite{xiang,blase} 
that supercells offer a more
realistic model than a virtual crystal treatment.  Our calculation of a
C$_{31}$B supercell indicated strong ordered-boron effects (which are
unphysical), and our CPA alloy calculations\cite{cpa} 
give bands like the virtual 
crystal model, with small disorder broadening added.
Another factor is anharmonicity,
which includes conventional anharmonicity  
and the non-adiabatic effects that cause the nonlinearity
of $\Delta E(u)/u^2$ in Fig. \ref{energy}.
Making the anharmonic corrections need not change the
effective phonon frequency greatly, as shown for MgB$_2$ by Lazzeri
{\it et al.}\cite{mauri} who found that for MgB$_2$ three- and 
four-phonon corrections gave strongly canceling corrections to the vibrational
frequency.   
The validity of the Migdal-Eliashberg itself
becomes an interesting question, and more so 
for lower doping levels.  For the 2.5\% concentration considered
here, the ratio of phonon frequency to electron energy scales 
is $\omega/E_F$ = 0.25,
certainly not the small parameter that is usually envisioned as a
perturbation expansion parameter.  Doped diamond provides a new system
in which to investigate non-adiabatic effects.

Now we summarize.  Based on the experimental
information available so far, the B doping
level in diamond achieved by Ekimov {\it et al.} should result in hole
doping of the diamond valence bands to a level E$_F \approx 0.6$ eV.  
Calculations bear out the analogy to MgB$_2$ that deformation potentials
due to bond stretching are extremely large, and 
evaluation of the hole-phonon coupling strength using conventional theory
leads to $\lambda \approx$ 0.55, a renormalization of the optic mode frequency 
by -20\%, and T$_c$ in the 5-10 K range.  
These results indicate that phonon coupling is the likely
candidate for the pairing mechanism, consistent with the conclusions of
Boeri {\it et al.}\cite{oka}
The low carrier density (for a metal) implies both poor screening of the
Coulomb interaction and the intrusion of non-adiabatic effects, 
which are primary candidates for further study.  Higher doping levels
should increase T$_c$, but probably not to anything like that occurring in 
MgB$_2$.  Definitive calculations will require CPA calculations of the 
electron-phonon coupling characteristics.

The authors acknowledge Z. Fisk for pointing out Ref.\cite{ekimov}, 
and J. Kune\v{s} and K. Koepernik for technical advice and
J. Kune\v{s} for a critical reading of the
manuscript.  This work was supported by National Science
Foundation Grant DMR-0114818.

\end{document}